\documentclass[12pt]{iopart}
\usepackage[round]{natbib}
\newcommand{\newblock}{}
\usepackage{lineno}
\modulolinenumbers[5]
\usepackage{datetime}
\newdateformat{monthyeardate}{\monthname[\THEMONTH], \THEYEAR}
\expandafter\let\csname equation*\endcsname\relax
\expandafter\let\csname endequation*\endcsname\relax
\usepackage{amsmath}
\usepackage{amsfonts}
\usepackage{graphicx}
\usepackage[textheight = 8.5in]{geometry}
\usepackage[dvipsnames]{xcolor}
\usepackage{booktabs}
\usepackage{siunitx}
\usepackage{multirow}
\usepackage{rotating}        
\usepackage{threeparttable}  
\usepackage{booktabs}         
\usepackage{adjustbox}
\usepackage{makecell}

\usepackage{soul}
\makeatletter
\renewcommand\st[1]{\@bsphack\@esphack}%
\makeatother

\newcommand{\ud}{\mbox{d}}

\begin{document}

\title[IMPULSED dMRI denoising]{Physics-Assisted Deep Learning Denoising for Stabilized IMPULSED dMRI Microenvironment Parameter Fitting}

\author{Wen Li$^{1}$, Yan Dai$^{2}$, Arely Perez Rodriguez$^{2}$, Todd Aguilera $^{2}$, Jie Deng$^{2}$, Xun Jia$^{1}$}

\address{$^1$Department of Radiation Oncology and Molecular Radiation Sciences, Johns Hopkins University, Baltimore, MD, USA\\$^2$Department of Radiation Oncology, University of Texas Southwestern Medical Center, Dallas, TX, USA\\

Email: xunjia@jhu.edu
}

\vspace{10pt}
\begin{indented}
\item[]\today
\end{indented}

\begin{abstract}
\\

\textit{Objective.} Diffusion-weighted MRI (dMRI) is a powerful tool for quantifying cellular microenvironment parameters. However, the inherently low signal-to-noise ratio (SNR) of dMRI can compromise the accuracy and reliability of parameter estimation. This study proposes a physics-assisted deep learning (DL)-based denoising framework designed to enhance dMRI signal quality and improve the robustness of subsequent biophysical model fitting, with potential relevance to low-SNR settings such as clinical 1.5 T MRI acquisitions.

\textit{Approach.} A dataset of paired noise-free and Rician-noise-corrupted dMRI signals was generated using the IMPULSED-dMRI signal model. Three denoising architectures were evaluated: Convolutional Neural Networks (CNN), Multilayer Perceptron (MLP), and Long Short-Term Memory (LSTM) networks. Denoised signals were then fitted to estimate cell diameter $d$, intracellular volume fraction $V_{\mathrm{in}}$, and extracellular apparent diffusion coefficient $D_\mathrm{ex}$. Performance was assessed using signal-domain Mean Absolute Error (MAE), fitted-parameter MAE, and fitting failure rate in synthetic IMPULSED-dMRI data with known ground-truth parameters and in \textit{in vitro} evaluation using HeLa and MC38 cell lines.

\textit{Main results.} DL-based processing substantially improved dMRI signal denoising. The MLP and LSTM achieved similar performance, with the LSTM slightly better overall, and both outperformed the CNN. Averaged across all signals, the LSTM reduced denoising MAE from 5.97\% to 1.58\%. In the subsequent model fitting step, the LSTM produced modest reductions in parameter MAE, from $5.34 \, \mu\text{m}$ to $4.02 \, \mu\text{m}$ for $d$, from 15.00\% to 11.96\% for $V_\mathrm{in}$, and from $0.77$ to $0.63 \, \mu\text{m}^2/\text{ms}$ for $D_\mathrm{ex}$. The dominant benefit was fitting stabilization, with the overall fitting failure rate reduced from 57.6\% to 17.7\%. In \textit{in vitro} experiments, relative to experimental references, the LSTM reduced MAE from $2.1 \, \mu\text{m}$ to $0.4 \, \mu\text{m}$ for $d$ and from 7.8 to 5.4 percentage points for $V_\mathrm{in}$, while reducing the mean overall fitting failure rate from 17.7\% to 0\%.

\textit{Significance.} The proposed framework improves dMRI signal quality and stabilizes subsequent IMPULSED-based microenvironmental parameter fitting. The primary value of this approach is improved fitting reliability under noisy dMRI conditions, with secondary gains in parameter accuracy, and it warrants further validation in heterogeneous tissues and \textit{in vivo} datasets.

\end{abstract}

%
\vspace{2pc}
\\

\noindent\submitto{\PMB}
%
\maketitle
%
%

\section{Introduction}

Characterizing the cellular microenvironment is essential for the early detection of therapeutic response in cancer. Parameters such as cell size and intracellular volume fraction serve as valuable biomarkers, reflecting treatment efficacy before anatomical changes become visible \citep{thoeny2010predicting}, thereby offering an important opportunity to personalize treatment strategies. While tissue biopsy remains the gold standard for obtaining this data, it is inherently invasive and carries risks of pain, bleeding, or infection \citep{loeb2013systematic}. Furthermore, due to intra-tumoral heterogeneity, localized tissue sampling may fail to provide a comprehensive characterization of the entire lesion. In some scenarios, biopsy may also be technically challenging or even prohibited depending on the anatomical locations and disease conditions \citep{hersh2020safety}. Hence, there is a critical need for noninvasive imaging techniques capable of comprehensive microenvironmental characterization.

Diffusion-weighted MRI (dMRI) has emerged as a powerful noninvasive tool for probing the tumor microenvironment. In dMRI, the measured signal reflects the microscopic motion of water molecules, which is influenced by features such as cell size, diffusion coefficients, and intracellular volume fraction. By applying biophysical signal models that link these microenvironmental properties to the observed signal data, it becomes possible to infer quantitative parameters of the tumor microenvironment from the measured data \citep{bammer2003basic}.

One important dMRI technique for estimating cellular microenvironment parameters is the IMPULSED (Imaging Microstructural Parameters Using Limited Spectrally Edited Diffusion) approach. \citet{jiang2016quantification} utilized temporal diffusion spectra obtained by combining pulsed gradient spin echo (PGSE) and oscillating gradient spin echo (OGSE) sequences to quantify cell diameter ($d$) and intracellular volume fraction ($V_{\mathrm{in}}$), and validated these estimates through in vitro experiments. The robustness of the method was further demonstrated in vivo across three types of human colon cancers \citep{jiang2017vivo}. \citet{drobnjak2016pgse} explored the optimal PGSE and OGSE settings for axon diameter estimation. Their results showed that low-frequency OGSE provides higher sensitivity in realistic conditions involving fibers with unknown or dispersed orientations. They also highlighted that MR systems equipped with stronger gradient hardware can further enhance sensitivity to axon diameter. More recently, \citet{wang2025microstructural} investigated the feasibility of using PGSE and OGSE to estimate cellular microenvironmental parameters in breast cancer patients and confirmed that the IMPULSED approach is a promising tool for characterizing breast cancer microstructural features and holds potential for prognostic risk assessment.

Despite the promising results, significant challenges remain due to the inherently low signal-to-noise ratio (SNR) of dMRI signals. These challenges may be especially relevant in low-SNR acquisition settings, including some clinical 1.5 T MRI workflows where scan time and sequence options are constrained \citep{arnold2023low,pogarell2024modern}. This limitation is partly related to main magnetic field strength ($B_0$), as the net magnetization vector is directly proportional to $B_0$ \citep{ladd2007high}. Signal averaging is a traditional method for improving SNR by repeating dMRI acquisitions \citep{madore1996new,eichner2015real}. However, the resulting increase in scan time may be prohibitive for clinical workflows. Beyond patient discomfort, longer scans are susceptible to motion-induced artifacts, which introduce registration errors between images at various $b$-values and bias the subsequent parameter estimation.

Post-processing through denoising offers a promising solution to these SNR limitations. Classical denoising techniques, such as Gaussian smoothing, Non-Local Means, and BM3D, rely on generalized mathematical assumptions like structural smoothness and self-similarity \citep{halfaoui2015improving,luo2012generalized,alnuaimy2024bm3d}. However, these approaches do not explicitly incorporate the IMPULSED signal model or acquisition-specific signal decay behavior, which may limit their ability to preserve model-relevant features in this application. In recent years, machine learning, and specifically deep learning (DL), has emerged as a new category of tools for signal denoising by learning noise patterns directly from data \citep{zhi2023coarse,li2022artificial,shen2020introduction,shan2020synergizing}. This data-driven nature allows models to account for realistic, scanner-specific noise, such as Rician distributions, and model specific signals, such as the IMPULSED model, making them highly adaptable to dMRI acquisitions for specific applications \citep{le2021deep,hu2025report}.

In this study, we present a physics-assisted DL framework for the denoising of IMPULSED-dMRI signals to improve the robustness of cellular microenvironment parameter estimation. We developed and compared multiple DL architectures to identify the optimal model for signal recovery, utilizing a physics-assisted data generation strategy where the models were trained on synthetic signal pairs derived directly from the IMPULSED biophysical equations. Importantly, we employed a two-step approach, in which the signals were first denoised via the DL model and subsequently fed into the biophysical model for parameter fitting. By using physics-assisted training to connect noisy IMPULSED-dMRI measurements with noise-free theoretical signal behavior, this methodology preserves explainability through a mathematically grounded link between the signal and the biological parameters. In contrast to direct parameter-mapping DL approaches that map raw signals directly to parameters \citep{barbieri2020deep,kaandorp2021improved,ottens2022deep}, our framework provides a practical and transparent pathway to reduce fitting instability while maintaining full biophysical interpretability.

\section{Methods}

\subsection{Data Preparation}
\label{sec:Data Preparation}
\subsubsection{dMRI Signal Model}
\label{sec:Signal Model}
In this study, we simulated dMRI signals based on real cell parameters and a dMRI imaging protocol using a well-established two-compartment IMPULSED model \citep{jiang2016quantification}. The efficacy of this model has been extensively demonstrated in previous studies \citep{alexander2010orientationally,xu2014mapping,assaf2004new,jiang2016quantification,li2014fast}. In this model, five cell microenvironment parameters were considered, including cell diameter ($d$), intracellular diffusion coefficient ($D_\mathrm{in}$), intracellular volume fraction ($V_\mathrm{in}$), extracellular diffusion coefficient ($D_\mathrm{ex}$), and the slope of extracellular diffusion coefficient with respect to oscillation frequency ($\beta_\mathrm{ex}$). Both pulsed gradient spin echo (PGSE) and oscillating gradient spin echo (OGSE) sequences are incorporated in the IMPULSED model. PGSE provides sensitivity to relatively large cellular dimensions through long diffusion times, whereas OGSE, especially at higher frequencies, enables probing of shorter diffusion times and thus improves sensitivity to small-scale structures. The dMRI signal $S$ at a given $b$ value can be expressed by:
\begin{equation}
    S(b) = V_\mathrm{in} \cdot S_\mathrm{in}(b) + (1 - V_\mathrm{in}) \cdot S_\mathrm{ex}(b),
    \label{Eq_S}
\end{equation}
where $S_\mathrm{in}(\cdot)$ and $S_\mathrm{ex}(\cdot)$ are intra- and extracellular signal components, and $V_\mathrm{in}$ is the intracellular water fraction. $b = \gamma^2 \int_0^{+\infty} \left|\int_0^t g(t'|G, \delta, \Delta)\,\ud t'\right|^2 \ud t$ is the diffusion weight, where $\gamma$ is the gyromagnetic ratio, $G$ is the diffusion gradient strength, $\delta$ is the gradient duration, and $\Delta$ represents the gradient separation. $g(\cdot)$ is the diffusion pulse's amplitude as a function of time, which depends on the specific pulse sequence.

The OGSE signal is measured using cosine-modulated gradient waveforms. The intracellular signal can be expressed as:
\begin{equation}
    \begin{aligned}
S_\mathrm{in} (b|\text{OGSE}) &= \exp \Bigg( 
-2 (\gamma G)^2 \sum_{k}
\frac{B_k \lambda_k^2 D_\mathrm{in}^2}{(\lambda_k^2 D_\mathrm{in}^2 + 4\pi^2 f^2)^2} \\
&\quad \times \Bigg\{ \frac{(\lambda_k^2 D_\mathrm{in}^2 + 4\pi^2 f^2)}{\lambda_k D_\mathrm{in}} 
\Bigg[ \frac{\delta}{2} + \frac{\sin(4\pi f \delta)}{8\pi f} \Bigg] \\
&\quad - 1 + \exp(-\lambda_k D_\mathrm{in} \delta) 
+ \exp(-\lambda_k D_\mathrm{in} \Delta) (1 - \cosh(\lambda_k D_\mathrm{in} \delta)) 
\Bigg\} \Bigg),
    \end{aligned}
    \label{Eq_Sin_OGSE}
\end{equation}
where $B_k$ and $\lambda_k$ are parameters that depend on the cell diameter. $f$ is the oscillation frequency. The extracellular OGSE signal is modeled as:
\begin{equation}
    \begin{aligned}
S_\mathrm{ex}(b|\text{OGSE}) = \exp\left[-b\left(D_\mathrm{ex} + \beta_\mathrm{ex} \cdot f\right)\right],
    \end{aligned}
    \label{Eq_Sex_OGSE}
\end{equation}
where $\beta_\mathrm{ex}$ denotes the slope of the extracellular free diffusion coefficient with respect to the oscillation frequency $f$, which indirectly reflects microenvironment properties.

When $f \to 0$, the cosine-modulated OGSE pulse degenerates into a PGSE pulse, and the intracellular and extracellular signals can be expressed as:
\begin{equation}
    \begin{split}
        S_\mathrm{in}(b|\text{PGSE}) &= \exp \Bigg( 
        -2 \left(\frac{\gamma G}{D_\mathrm{in}}\right)^2 
        \sum_{k} \frac{B_k}{\lambda_k^2} 
        \Bigg[ \lambda_k D_\mathrm{in} \delta - 1 
        + \exp(-\lambda_k D_\mathrm{in} \delta) \\
        &\quad + \exp(-\lambda_k D_\mathrm{in} \Delta) 
        \left(1 - \cosh(\lambda_k D_\mathrm{in} \delta) \right) 
        \Bigg] \Bigg),\\
        S_\mathrm{ex}(b|\text{PGSE}) &= \exp(-bD_\mathrm{ex})
    \end{split}
    \label{Eq_S_PGSE}
\end{equation}

\subsubsection{Data generation}
\label{sec:Signal Simulation}
In this study, we focused on the following ranges of the five cell microenvironment parameters: $d$ = 4 - 28 $\mu$m, $D_\mathrm{in}$ = 0.2 - 3.38 $\mu$m$^2$/ms, $V_\mathrm{in}$ = 0\% - 100\%, $D_\mathrm{ex}$ = 0.2 - 3.38 $\mu$m$^2$/ms, and $\beta_\mathrm{ex}$ = 0 - 10 $\mu$m$^2$, representing physiologically realistic human microenvironment characteristics \citep{xu2021mri}. A clinical IMPULSED-dMRI acquisition protocol was employed, including one PGSE sequence and two OGSE sequences (OGSEn1 and OGSEn2) with cycle number N = 1 and N = 2, respectively. Major parameters for PGSE included $\delta=21$ ms, $\Delta=68$ ms, $t_\mathrm{rise}=1.355$ ms, $b$-values = (0, 10, 100, 200, 300, 400, 500, 600, 700, 800, 900, 1000) s/mm$^2$. For OGSEn1, the parameters were  $\delta=60$ ms, $\Delta=68$ ms, $t_\mathrm{rise}=1.32$ ms, $b$-values = (0, 10, 100, 200, 300, 400, 500, 600, 700, 800, 900, 1000) s/mm$^2$, and those for OGSEn2 included $\delta=60$ ms, $\Delta=68$ ms, $t_\mathrm{rise}=1.32$ ms, $b$-values = (0, 10, 35, 70, 105, 140, 175, 210, 245, 280, 315, 350) s/mm$^2$. 

To facilitate the physics-assisted DL framework, a comprehensive dataset of paired noisy and noise-free dMRI signals was generated. First, 100,000 sets of microenvironment parameters ($d$, $V_{\mathrm{in}}$, $D_{\mathrm{in}}$, $D_{\mathrm{ex}}$, and $\beta_{\mathrm{ex}}$) were randomly sampled in the parameter ranges to ensure uniform and efficient coverage of the IMPULSED signal space. Each parameter was independently sampled from a uniform distribution over the specified range. These sampled parameter sets were used as inputs for the IMPULSED biophysical equations (Eqs. (\ref{Eq_S})-(\ref{Eq_S_PGSE})) to calculate the corresponding noise-free dMRI signals for the specific PGSE and OGSE acquisition settings. To approximate magnitude-MRI noise in a controlled simulation setting, Rician noise was then added to each noise-free signal to generate its paired noisy counterpart. Specifically, the noisy signal was generated as $S_{\mathrm{noisy}}(b) = |[S(b)\cos(\phi)+n_r]+\mathrm{i}[S(b)\sin(\phi)+n_i]|$, where $n_r$ and $n_i$ are independent zero-mean Gaussian noise components with standard deviation $\sigma$, and $\phi$ is a phase factor randomly sampled in $[0,2\pi)$. The corresponding SNR was defined as $S(0)/\sigma = 1/\sigma$ \citep{gudbjartsson1995rician}. The model-development dataset included noise levels $\sigma$ in 0.01 - 0.12, corresponding to SNR ranging approximately from 100 down to 8. This range was selected to span different noise levels that may occur in real dMRI acquisitions and to evaluate the robustness of the denoising model under varying image-quality conditions.

\subsection{Denoising Model Development}
\label{sec:Model development}

We decoupled the estimation of cell microenvironment parameters into two steps: denoising and fitting. First, we trained a DL model to map noisy dMRI signals to their corresponding noise-free counterparts. The denoised signals were then fitted to the cell parameters using the established biophysical model. We adopted this two-step strategy rather than directly feeding noisy dMRI signals into machine learning models for parameter estimation because the latter requires the model to simultaneously suppress noise and infer parameters, which increases task complexity and the risk of overfitting to the noise distribution. In contrast, the denoising-fitting approach leverages the data-driven approach solely for noise suppression, leaving parameter estimation to the biophysical model fitting step. This separation supports stable fitting while maintaining consistency with the IMPULSED-dMRI signal model and enhancing interpretability.

\subsubsection{Denoising Models}

\begin{figure}
  \centering
  \includegraphics[width=1\textwidth]{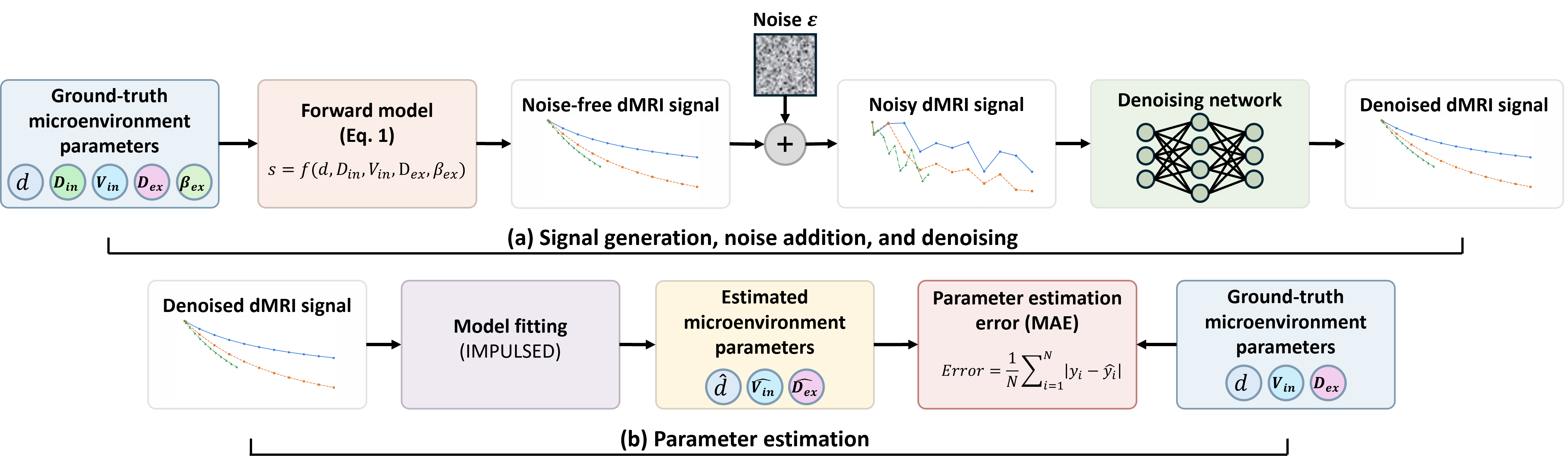}
  \caption{Overall workflow of this study. (a) Workflow of signal generation, noise addition, and denoising; (b) Workflow of parameter estimation.}
  \label{fig:workflow}
\end{figure}

\begin{figure}
  \centering
  \includegraphics[width=1\textwidth]{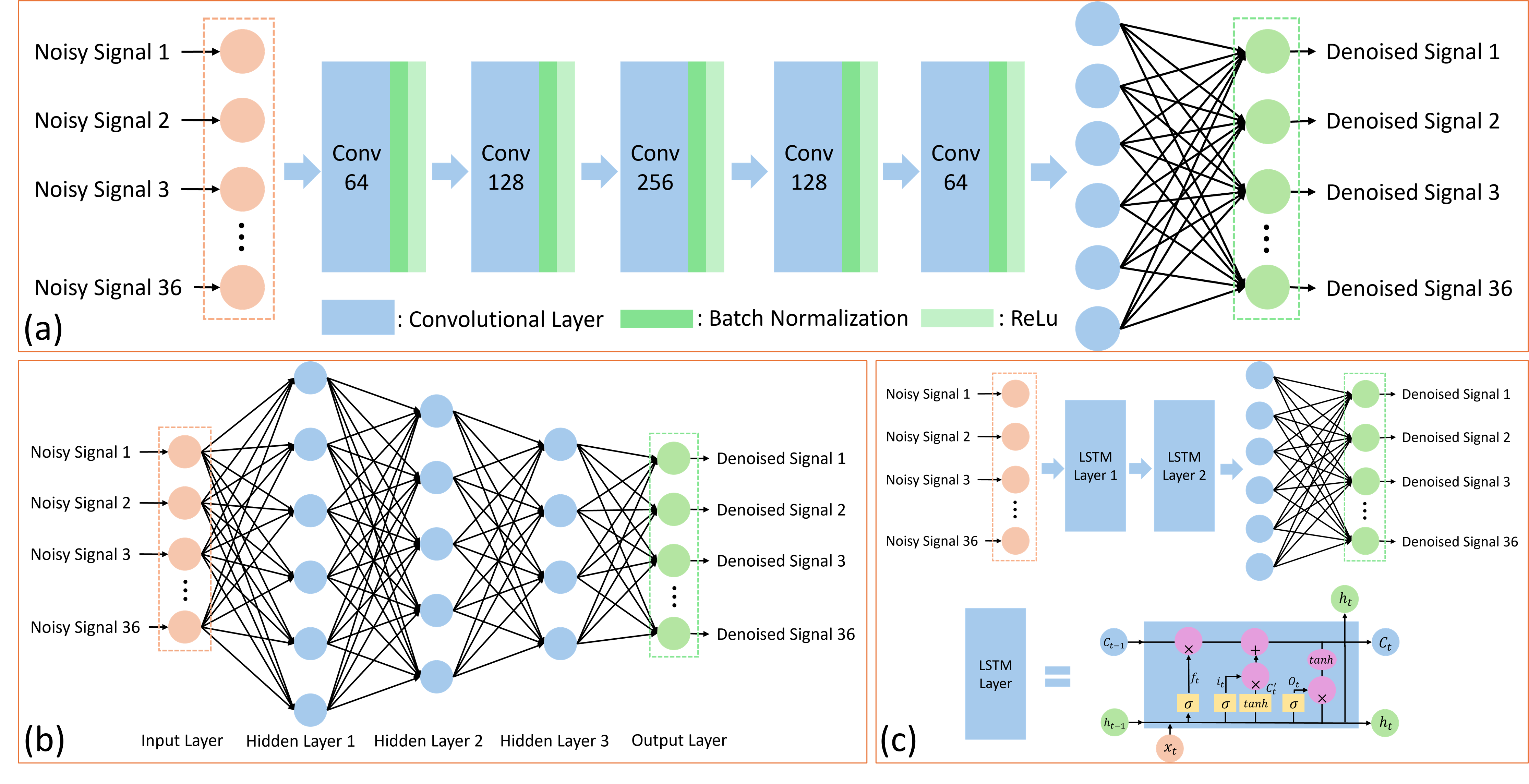}
  \caption{Illustration of DL models used in this study. (a) Convolutional Neural Network (CNN); (b) Multilayer Perceptron (MLP); (c) Long Short-Term Memory (LSTM).}
  \label{fig:3DLModels}
\end{figure}

The overall workflow of the proposed denoising framework and model architectures are illustrated in Fig.~\ref{fig:workflow} and Fig.~\ref{fig:3DLModels}. Each model is designed to map a sequence of 36 1D noisy measurements acquired at various $b$-values to their corresponding noise-free counterparts. The 36-point input was formed by concatenating the PGSE, OGSEn1, and OGSEn2 signal vectors in the order presented in Section~\ref{sec:Signal Simulation}, with the $b$-values ordered within each sequence. Formally, we denote the denoising operation as $y = f(x \mid \theta)$,
where $x \in \mathbb{R}^{36}$ represents the noisy dMRI signal input, $y \in \mathbb{R}^{36}$ is the predicted denoised signal, and $\theta$ denotes the trainable network parameters. 

To identify an effective architecture for this task, we evaluated three distinct 1D DL models including (i) a 1D convolutional neural network (CNN) \citep{kiranyaz20211d}, (ii) a multilayer perceptron (MLP) \citep{popescu2009multilayer}, and (iii) a long short-term memory (LSTM) network \citep{yu2019review}. 

\textbf{Convolutional Neural Network (CNN):} Given the relatively low complexity of the dMRI signal, we applied a shallow 5-layer CNN to prevent overfitting. Each convolutional layer used a kernel size of 5 with zero padding, and the output channels were 64, 128, 256, 128, and 64, respectively. A fully connected layer was added at the end. Each convolutional layer was followed by batch normalization and ReLU activation. A Sigmoid activation function was applied to the output layer to constrain the predictions to the range of 0 to 1 and prevent out-of-range outputs.

\textbf{Multilayer Perceptron (MLP):} The MLP is a feedforward deep neural network composed of fully connected layers with nonlinear activation functions. In our study, we implemented a 4-layer MLP with output dimensions of 128, 64, 32, and 36, respectively. ReLU was used as the activation function after each layer and a Sigmoid activation function was applied to the output layer to constrain the predictions to the range of 0 to 1 and prevent out-of-range outputs. 

\textbf{Long Short-Term Memory Network (LSTM):} LSTM network is a type of recurrent neural networks capable of modeling long-term dependencies in sequential data, making them suitable for dMRI denoising given the monotonic decay of signal with increasing $b$-values. In this study, although the input is not temporal, the ordered concatenation of PGSE, OGSEn1, and OGSEn2 signals contains structured signal decay patterns that can be modeled using recurrent units. Each LSTM unit includes an input gate (controls new information stored in the cell state), a forget gate (determines which information from the previous state to discard), and an output gate (controls information passed to the next layer) \citep{sherstinsky2020fundamentals}. We implemented a 2-layer LSTM with 128 and 16 output channels, followed by a fully connected layer applied to the hidden state at the final time step to generate the output. A Sigmoid activation function was applied to the output layer to constrain the predictions to the range of 0 to 1 and prevent out-of-range outputs. 

\subsubsection{Denoising Model development}

The network parameters $\theta$ were optimized using a physics-assisted training strategy by minimizing the L1 loss function, defined as $
    \mathcal{L}(\theta) = \frac{1}{N}\sum_{i=1}^{N} |y_i - f(x_i \mid \theta)|$,
where $x_i$ is the noisy dMRI signal input, $y_i$ represents the corresponding ground-truth noise-free signal generated by the IMPULSED equations, and $N$ is the number of samples. We chose the L1 loss over the L2 loss, because it is less sensitive to outliers.

The synthetic dataset of 100,000 signal pairs was randomly partitioned into training, validation, and testing subsets using a 70:15:15 ratio. To ensure a fair comparison, all three architectures (CNN, MLP, and LSTM) were trained using a consistent empirical configuration: the Adam optimizer with a learning rate of 0.001, up to 500 epochs to encourage convergence, and a batch size of 128 per iteration. The final model was selected based on the lowest validation loss function value; convergence was confirmed by validation loss plateauing. Although the denoising models took 36-dimensional signals as both input and output, only 33 dimensions were effectively denoised. This is because after normalization, the first signal in each of the three sequences was fixed at 1 by definition. Therefore, these three $b=0$ entries were excluded from the loss and reset to 1 in the output.

The training was implemented in Python using the PyTorch framework, and the hardware utilized for all experiments was a Windows 11 workstation (Version: 23H2) equipped with an NVIDIA A4000 GPU (16 GB memory) and an Intel Xeon Silver 4214R CPU (2.40 GHz).

\subsection{Data Fitting}
After denoising, the dMRI signals were fitted to the biophysical model (Section ~\ref{sec:Signal Model}) to estimate the cell microenvironment parameters. Prior IMPULSED sensitivity analyses have shown that $D_\mathrm{in}$ and $\beta_\mathrm{ex}$ are poorly conditioned under this acquisition protocol and can reduce fitting robustness when estimated simultaneously with the other parameters \citep{Li2026Investigating}. Therefore, consistent with that prior analysis, this study focused the fitting step on $d$, $V_\mathrm{in}$, and $D_\mathrm{ex}$, while keeping $D_\mathrm{in}$ and $\beta_\mathrm{ex}$ fixed. 

Curve fitting was performed using nonlinear least-squares optimization, where the objective was to minimize the squared error between the denoised signals and the model-predicted signals across all $b$-values and sequences. During fitting, $D_\mathrm{in}$ and $\beta_\mathrm{ex}$ were fixed at 1.56 $\mu$m$^2$/ms and 0 $\mu$m$^2$, respectively. Parameter bounds were constrained to physiologically realistic ranges ($d$ = 4 - 28 $\mu$m, $V_\mathrm{in}$ = 0\% - 100\%, and $D_\mathrm{ex}$ = 0.2 - 3.38 $\mu$m$^2$/ms), and initial values were set as $d$ = 18 $\mu$m, $V_\mathrm{in}$ = 60\%, and $D_\mathrm{ex}$ = 1.5 $\mu$m$^2$/ms. Optimization was carried out using the Trust Region Reflective algorithm with default convergence tolerance of $10^{-8}$. The maximum number of function evaluations was set to 10000 to ensure convergence of the nonlinear least-squares fitting. A fit was classified as failed if any estimated parameter was at its lower or upper bound, because boundary solutions indicate that the optimizer was driven to the imposed constraints rather than to a stable interior solution. The overall fitting failure rate was defined as the percentage of fits in which at least one estimated parameter reached its lower or upper bound. To characterize the failure modes, we further reported parameter-specific failure rates for $d$, $V_\mathrm{in}$, and $D_\mathrm{ex}$, defined as the percentage of fits in which each individual parameter reached a boundary. These parameter-specific rates were calculated separately and were not mutually exclusive, because a single failed fit could involve more than one parameter reaching its boundary.

\subsection{Evaluations}
\label{sec:Evaluation}

\subsubsection{Evaluations using simulated data}

To comprehensively evaluate the developed models, we assessed performance across the signal domain to measure denoising fidelity, and then the parameter domain to determine the impact of denoising on biophysical model fitting.

For the evaluation in the signal domain, beyond assessing general performance on the independent test set (noise levels $\sigma \in [0.01, 0.12]$), we evaluated model robustness across a discrete range of SNRs. Specifically, Rician noise was added to the noise-free testing signals to simulate SNR values of 8, 10, 15, 20, 30, 40, 60, 80, and 100 (corresponding to noise levels of 0.125, 0.10, 0.067, 0.05, 0.033, 0.025, 0.017, 0.013, and 0.01, respectively). We utilized a denser sampling of points at lower SNRs, as denoising architectures are theoretically more sensitive to high-noise regimes \citep{kaandorp2021improved}. Denoising accuracy in the signal domain was quantified using the Mean Absolute Error (MAE) between the denoised and ground-truth noise-free signals.

For the evaluation in the fitted parameter domain, we performed biophysical model fitting on both the original noisy signals and the denoised signals produced by the MLP, CNN, and LSTM architectures. This allowed for a direct comparison of how each denoising approach improved the estimation of cellular parameters ($d$, $V_\mathrm{in}$, and $D_\mathrm{ex}$). This assessment was conducted under both the randomly sampled noise distribution and the predefined discrete SNR levels mentioned above. The accuracy of the parameter estimation was quantified by calculating the MAE between the parameters derived from the fitting process and the known ground-truth values used to generate the synthetic data. We also measured the overall fitting failure rate as the main robustness metric and parameter-specific failure rates to identify which fitted variables most frequently contributed to unstable boundary solutions across different scenarios. 

\subsubsection{Evaluations using \textit{in vitro} experimental data}
An \textit{in vitro} experiment was conducted using two cell types:  HeLa and MC38 cell lines. HeLa is the oldest human cell line, derived from cervical cancer cells. It is immortalized, easy to culture, and widely used in biomedical research \citep{masters2002hela}. The MC38 cell line is a murine colon adenocarcinoma model derived from a C57BL/6 mouse, commonly used in colorectal cancer studies and immunotherapy research \citep{hos2020identification}.

These cells were cultured in complete DMEM supplemented with 10\% Fetal Clone II serum (FC II) and 1$\times$ Penicillin-Streptomycin under standard culture conditions. Cells were passaged at approximately 80 - 90\% confluency. For harvesting, cells were dissociated using 0.25\% trypsin-EDTA, neutralized with complete medium, pelleted, and counted using the TC20\textsuperscript{TM} automated cell counter (Bio-Rad Laboratories, Inc. Hercules, CA). For fixation, $1\times 10^8$ MC38 or HeLa cells per sample were resuspended in 4\% paraformaldehyde (PFA) in PBS and incubated for 1 hour on ice. After fixation, cells were washed once with PBS and transferred to 1.5 mL microcentrifuge tubes. Samples were centrifuged at 3,000$\times$g for 2 minutes at 4 $^\circ$C, and supernatants were carefully aspirated to avoid disturbing the pellets. An additional MC38 sample was centrifuged at 6,000$\times$g to evaluate pellet compaction effects and its impact on cell microenvironment parameter estimation. Pellets remained intact throughout all processing steps. Following preparation, the tubes containing the fixed cell pellets were immersed in Fomblin during the dMRI acquisition. 

All three tubes were scanned on a clinical 1.5 T MRI system (Ingenia Ambition X, Philips Healthcare, Netherlands). The PGSE, OGSEn1, and OGSEn2 imaging protocols followed those described in Section~\ref{sec:Signal Simulation}. Additional imaging parameters included TR = 3000 ms, TE = 139 ms, voxel size = $2\times 2\times 5$ mm$^3$, FOV = $160 \times 160$ mm$^2$, 4 slices, SENSE factor = 3, and half-scan factor = 0.63. The total scan time was 2 minutes 24 seconds for each sequence, with 12 seconds per $b$ value acquisition.

To extract the cell-containing voxels from the dMRI images, masks were manually generated and applied based on the PGSE $b$0 image by contouring the bright regions corresponding to cell pellets. After voxel selection, model fitting was first performed directly on the raw dMRI signals from the three tubes (i.e., noisy signals), and these results served as the baseline. The same dMRI signals were then denoised using the three trained models described in Section~\ref{sec:Model development}, and the denoised signals were used for microenvironment parameter fitting. Fitting results were compared against experimentally derived reference values for cell diameter $d$ and intracellular volume fraction $V_\mathrm{in}$. Specifically, the reference $d$ and cell count were measured using the TC20\textsuperscript{TM} automated cell counter, and the reference $V_\mathrm{in}$ was estimated as the total intracellular volume divided by the remaining sample volume after centrifugation:
$V_\mathrm{in} = \frac{N \pi d^3/6}{V_{\mathrm{pellet}}}$, where $N$ is the number of cells, $d$ is the cell diameter, and $V_{\mathrm{pellet}}$ is the remaining volume after removal of the supernatant. This calculation treats the cells as approximately spherical and uses the pellet volume as the total sample volume for the cell-containing compartment. Therefore, the resulting $d$ and $V_\mathrm{in}$ values were used as experimental references rather than absolute ground truth, because they may be affected by cell-size measurement uncertainty, fixation-related changes, pellet-volume measurement uncertainty, and residual extracellular space within the pellet. Experimental MAE was calculated as the mean absolute difference between the fitted values and the corresponding experimental references across the three cell-line datasets, and the mean overall fitting failure rate was calculated by averaging the overall fitting failure rates across HeLa, MC38-3, and MC38-6. The extracellular diffusion coefficient $D_\mathrm{ex}$ was not available in the experiments because no reliable experimental reference measurement was available; therefore, only the estimated values are reported in this study.


\section{Results}
\subsection{dMRI Signal Denoising Results}
\label{sec:denoising results}

\begin{figure}
  \centering
  \includegraphics[width=1\textwidth]{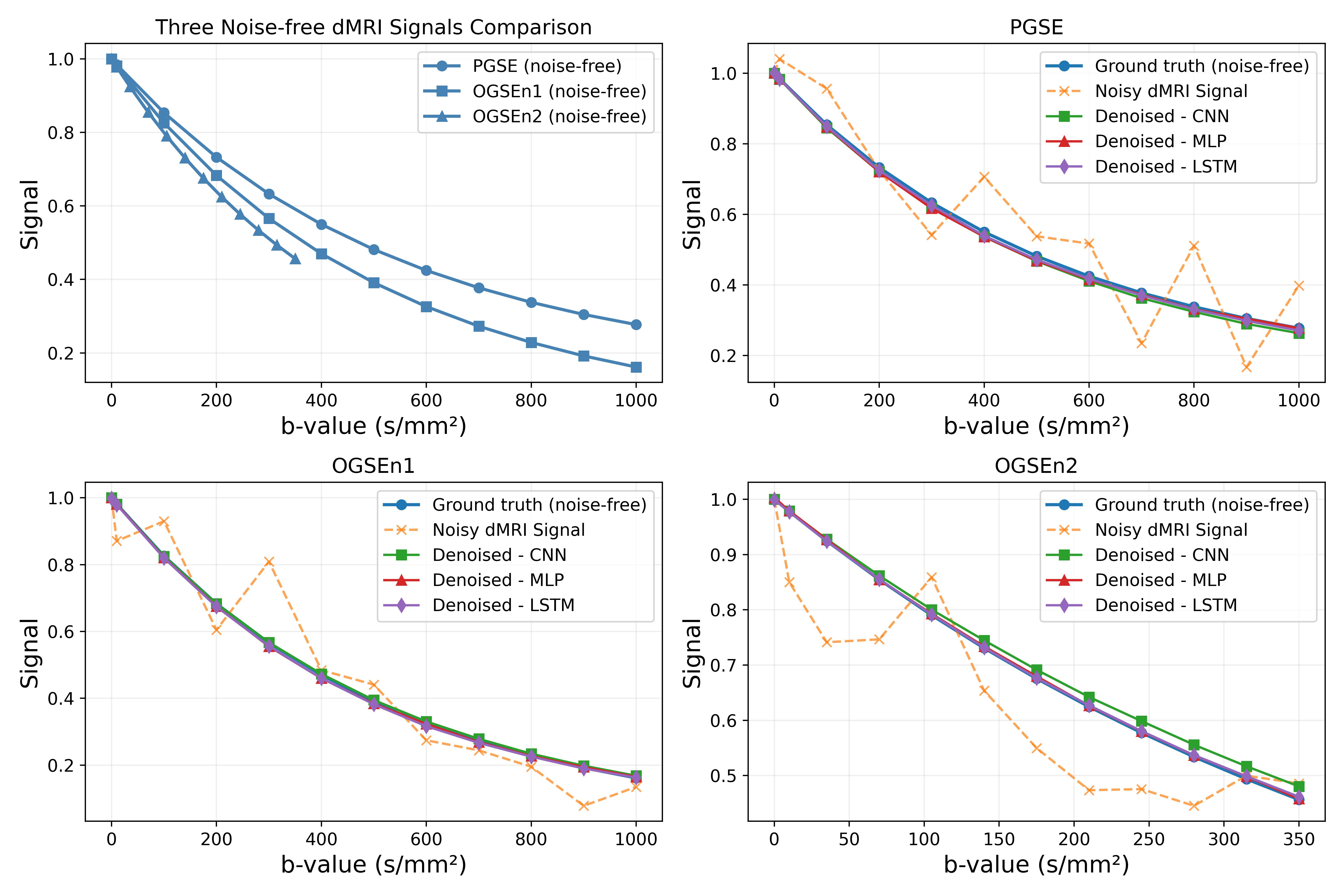}
  \caption{Denoising performance comparison among the three denoising models for each sequence for a representative test case.}
  \label{fig:denoising_results_main}
\end{figure}

Figure~\ref{fig:denoising_results_main} illustrates the noise-free dMRI signals for the PGSE, OGSEn1, and OGSEn2 sequences, along with the corresponding denoising results for each of the three models for a representative test case. As shown, the raw noisy signals deviate substantially from the noise-free reference; however, the DL models demonstrate the ability to recover the underlying signal, effectively learning to project noise-corrupted measurements back into the true signal manifold defined by the underlying dMRI signal model. Among the models, the CNN underperformed relative to the MLP and LSTM, particularly for the OGSEn2 sequence beyond the fourth signal point. This performance gap may be attributed to the relatively narrow dynamic range and subtle signal variations inherent to the OGSEn2 sequence (as seen in the top-left subfigure). These subtle features are more susceptible to being masked by noise, making it difficult for the CNN's convolutional kernels to distinguish between noise patterns and meaningful signal changes. In contrast, the LSTM and MLP architectures exhibit high fidelity in signal recovery, consistently tracking the ground-truth decay across all three acquisition protocols.

Table \ref{tab:mae_all_simplified} presents the MAE values for dMRI signal denoising using the three models over all cases. The noisy-signal baseline MAE was computed between the noise-free dMRI signal and signals corrupted with Rician noise. The top section of the table presents MAEs for each sequence, with each row corresponding to a single signal point associated with a specific $b$-value as presented in Section \ref{sec:Signal Simulation}. The lower section of the table presents the MAEs for each denoising model averaged over different sequences and $b$-values. Across all sequences and architectures, MAE exhibited a clear positive correlation with increasing $b$-values. This trend is consistent with the physical principles of dMRI, where stronger diffusion weighting results in exponential signal decay. The resulting decrease in signal magnitude leads to a substantial drop in the local SNR, rendering accurate signal recovery more challenging at high $b$-values.

\begin{table}[htbp]
\centering
\scriptsize
\caption{MAE ($\%$) Comparison across models and sequences. The upper section presents sequence-based results, while the lower section presents results across all signals. The first column lists the corresponding $b$-values in s/mm$^2$; when two values are shown, they indicate PGSE/OGSEn1 and OGSEn2, respectively. Noisy = noisy-signal baseline; Seq. Avg. = sequence-based average; All Sig. Avg. = all-signal-based average.}
\label{tab:mae_all_simplified}
\vspace{4pt}
\setlength{\tabcolsep}{4pt}
\renewcommand{\arraystretch}{1.1}
\begin{tabular}{@{}c @{\hspace{12pt}} cccc @{\hspace{12pt}} cccc @{\hspace{12pt}} cccc @{}}
\toprule
& \multicolumn{4}{c}{\textbf{PGSE}} & 
\multicolumn{4}{c}{\textbf{OGSEn1}} & 
\multicolumn{4}{c}{\textbf{OGSEn2}} \\
\cmidrule(r){2-5} \cmidrule(lr){6-9} \cmidrule(l){10-13}
\multirow{2}{*}{\makecell{$b$ value\\(s/mm$^2$)}} & 
\multirow{2}{*}{Noisy} & \multicolumn{3}{c}{Models} &
\multirow{2}{*}{Noisy} & \multicolumn{3}{c}{Models} &
\multirow{2}{*}{Noisy} & \multicolumn{3}{c}{Models} \\
\cmidrule(r){3-5} \cmidrule(lr){7-9} \cmidrule(l){11-13}
& & CNN & MLP & LSTM & & CNN & MLP & LSTM & & CNN & MLP & LSTM \\
\midrule
0  & 0.00 & 0.00 & 0.00 & 0.00 & 0.00 & 0.00 & 0.00 & 0.00 & 0.00 & 0.00 & 0.00 & 0.00 \\
10  & 7.40 & 0.14 & 0.14 & 0.18 & 7.36 & 0.13 & 0.18 & 0.19 & 7.35 & 0.16 & 0.17 & 0.19 \\
100; 35  & 7.08 & 1.17 & 0.95 & 0.97 & 6.95 & 1.04 & 0.83 & 0.84 & 7.21 & 0.50 & 0.43 & 0.43 \\
200; 70  & 6.77 & 1.94 & 1.57 & 1.57 & 6.67 & 1.68 & 1.30 & 1.31 & 7.07 & 0.93 & 0.77 & 0.75 \\
300; 105  & 6.66 & 2.44 & 1.96 & 1.94 & 6.49 & 2.08 & 1.61 & 1.60 & 6.94 & 1.30 & 1.06 & 1.04 \\
400; 140  & 6.51 & 2.76 & 2.22 & 2.18 & 6.29 & 2.32 & 1.79 & 1.77 & 6.76 & 1.63 & 1.33 & 1.29 \\
500; 175  & 6.39 & 2.97 & 2.38 & 2.34 & 6.15 & 2.48 & 1.91 & 1.89 & 6.65 & 1.91 & 1.56 & 1.51 \\
600; 210  & 6.31 & 3.12 & 2.49 & 2.47 & 6.05 & 2.61 & 2.04 & 2.00 & 6.57 & 2.17 & 1.75 & 1.71 \\
700; 245  & 6.18 & 3.24 & 2.60 & 2.58 & 5.96 & 2.75 & 2.15 & 2.13 & 6.48 & 2.38 & 1.93 & 1.88 \\
800; 280 & 6.18 & 3.35 & 2.70 & 2.68 & 5.86 & 2.89 & 2.30 & 2.28 & 6.33 & 2.57 & 2.10 & 2.03 \\
900; 315 & 6.04 & 3.47 & 2.82 & 2.80 & 5.82 & 3.05 & 2.46 & 2.45 & 6.33 & 2.74 & 2.22 & 2.18 \\
1000; 350 & 6.04 & 3.60 & 2.92 & 2.94 & 5.73 & 3.23 & 2.64 & 2.63 & 6.26 & 2.89 & 2.35 & 2.30 \\
\midrule
Seq. Avg. & 5.96 & 2.35 & 1.90 & \textbf{1.89} & 5.78 & 2.02 & 1.60 & \textbf{1.59} & 6.16 & 1.60 & 1.31 & \textbf{1.28} \\
\bottomrule
\toprule
\multirow{2}{*}{All Sig. Avg.} & \multicolumn{2}{c}{Noisy} & 
\multicolumn{2}{c}{CNN} & \multicolumn{2}{c}{MLP} & 
\multicolumn{2}{c}{LSTM} \\
 & \multicolumn{2}{c}{5.97} & 
\multicolumn{2}{c}{1.99} & \multicolumn{2}{c}{1.60} & 
\multicolumn{2}{c}{\textbf{1.58}} \\
\bottomrule
\end{tabular}
\scriptsize
\end{table}

Among the three models, the CNN had the highest residual errors, whereas the MLP and LSTM both achieved low denoising MAEs. The LSTM achieved the lowest sequence-averaged MAE across all three sequences ($1.89\%$, $1.59\%$, and $1.28\%$), although its performance was close to that of the MLP. Compared to the noisy-signal baseline MAE values of $5.96\%$, $5.78\%$, and $6.16\%$, the LSTM substantially reduced the error across the entire acquisition sequences. Averaged across all sequences, it reduced the aggregate MAE from a noisy-signal baseline value of $5.97\%$ to $1.58\%$.

\begin{figure}
  \centering
  \includegraphics[width=0.7\textwidth]{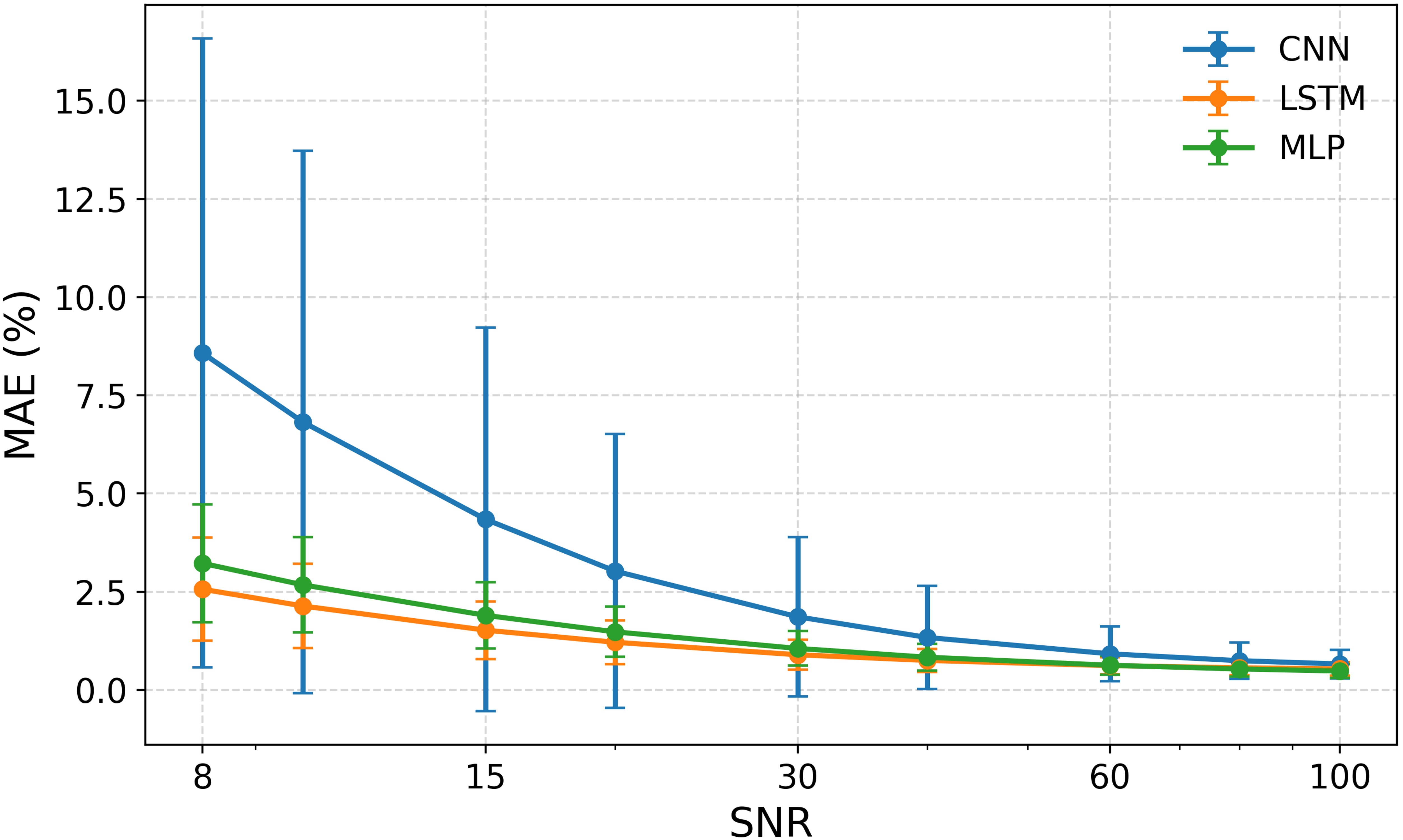}
  \caption{Comparison of dMRI denoising performance across SNR =  8-100. A logarithmic scale was applied to the SNR axis to improve visualization of low-SNR range.}
  \label{fig:denoising_results_models}
\end{figure}

Figure~\ref{fig:denoising_results_models} presents the denoising performance as a function of noise level for different denoising models. 
As expected, the MAE for all three architectures decreased monotonically as the SNR increased. The figure further shows that both the LSTM and MLP models consistently outperformed the CNN, yielding lower MAE and narrower standard deviations (STD). In the high-SNR regime (SNR $>$ 60), the performance of the three models converged, as the reduced noise contamination and improved signal fidelity naturally facilitate more accurate signal recovery across all architectures. This performance gap was particularly pronounced in the low-SNR regime (SNR $<$ 30), where the LSTM further demonstrated a slight advantage over the MLP in both accuracy in terms of MAE and precision in terms of STD. 

\subsection{Data Fitting Results}
\label{sec:fitting_results}

Following signal denoising, the dMRI data were utilized to estimate the three cell microenvironment parameters ($d$, $V_\mathrm{in}$, and $D_\mathrm{ex}$) across the testing dataset. Fitting the raw noisy signals yielded MAEs of 5.3 $\mu$m for cell diameter, 15.0\% for $V_\mathrm{in}$, and 0.8 $\mu$m$^2$/ms for $D_\mathrm{ex}$. The integration of DL-based denoising produced modest reductions in estimation errors across all three parameters. Consistent with the signal-domain results, the MLP and LSTM both improved parameter estimation, with the LSTM producing the lowest overall MAEs of 4.0 $\mu$m, 12.0\%, and 0.6 $\mu$m$^2$/ms for $d$, $V_\mathrm{in}$, and $D_\mathrm{ex}$, respectively. However, the main benefit of denoising was not the absolute reduction in MAE but the stabilization of the nonlinear fitting process: LSTM-based denoising reduced the parameter-specific failure rates from 26.5\% to 2.1\% for $d$, 8.8\% to 0.1\% for $V_\mathrm{in}$, and 33.9\% to 15.8\% for $D_\mathrm{ex}$, with the overall fitting failure rate reduced from 57.6\% to 17.7\%. Because the parameter-specific rates were computed separately for each parameter, they describe overlapping failure modes rather than mutually exclusive categories.

\begin{figure}
  \centering
  \includegraphics[width=1\textwidth]{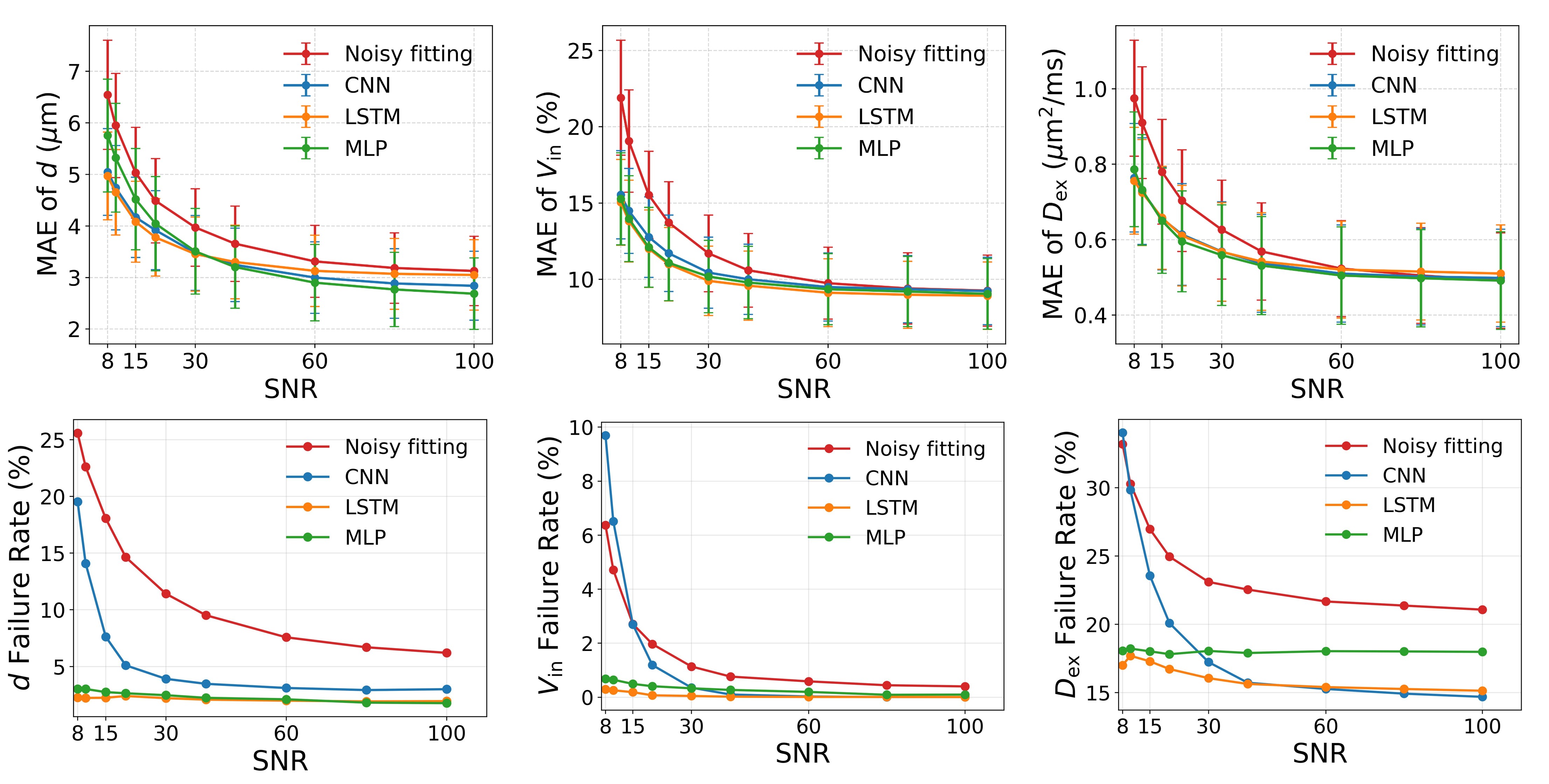}
  \caption{Comparison of fitting results from noisy and denoised dMRI signals across noise levels. The top row shows parameter MAE for $d$, $V_\mathrm{in}$, and $D_\mathrm{ex}$, and the bottom row shows the corresponding parameter-specific failure rates. A logarithmic scale was applied to the SNR axis to improve visualization of the low-SNR region.}
  \label{fig:denoising_results}
\end{figure}

Figure~\ref{fig:denoising_results} illustrates the estimation errors and parameter-specific failure rates for $d$, $V_\mathrm{in}$, and $D_\mathrm{ex}$ as a function of SNR. Across all three parameters, biophysical model fitting performed on denoised dMRI signals produced lower MAE values than direct fitting of raw noisy signals, although the magnitude of the accuracy improvement was modest relative to the reduction in fitting failures. Mirroring the signal-domain results, the MAE for all three models and direct fitting decreased monotonically as the SNR increased. Specifically, for the estimation of cell diameter $d$, LSTM-denoised signals achieved the highest accuracy at low SNRs (SNR $<$ 30), whereas the MLP yielded slightly lower MAE values when the SNR exceeded 30. For $V_\mathrm{in}$ and $D_\mathrm{ex}$ estimation, the LSTM and MLP demonstrated comparable performance across the entire SNR spectrum. As the SNR surpassed 60, the estimation errors of all four methods converged, suggesting that the marginal benefit of DL-based denoising diminishes in high-SNR regimes. 

In addition to modestly improving estimation accuracy, the DL-based denoising models reduced the parameter-specific failure rates across different SNR levels, especially the LSTM model. Boundary hits were more frequent at low SNR levels, where noise contamination can distort the signal and drive the fitting process toward parameter limits. Because multiple parameters could reach boundaries in the same failed fit, the parameter-specific rates were allowed to overlap. By suppressing noise while recovering the underlying signal pattern, the DL-based denoising models improved the stability of the fitting procedure and reduced the overall fitting failure rate.

\subsection{\textit{In Vitro} Evaluation Results}

\begin{figure}
  \centering
  \includegraphics[width=1\textwidth]{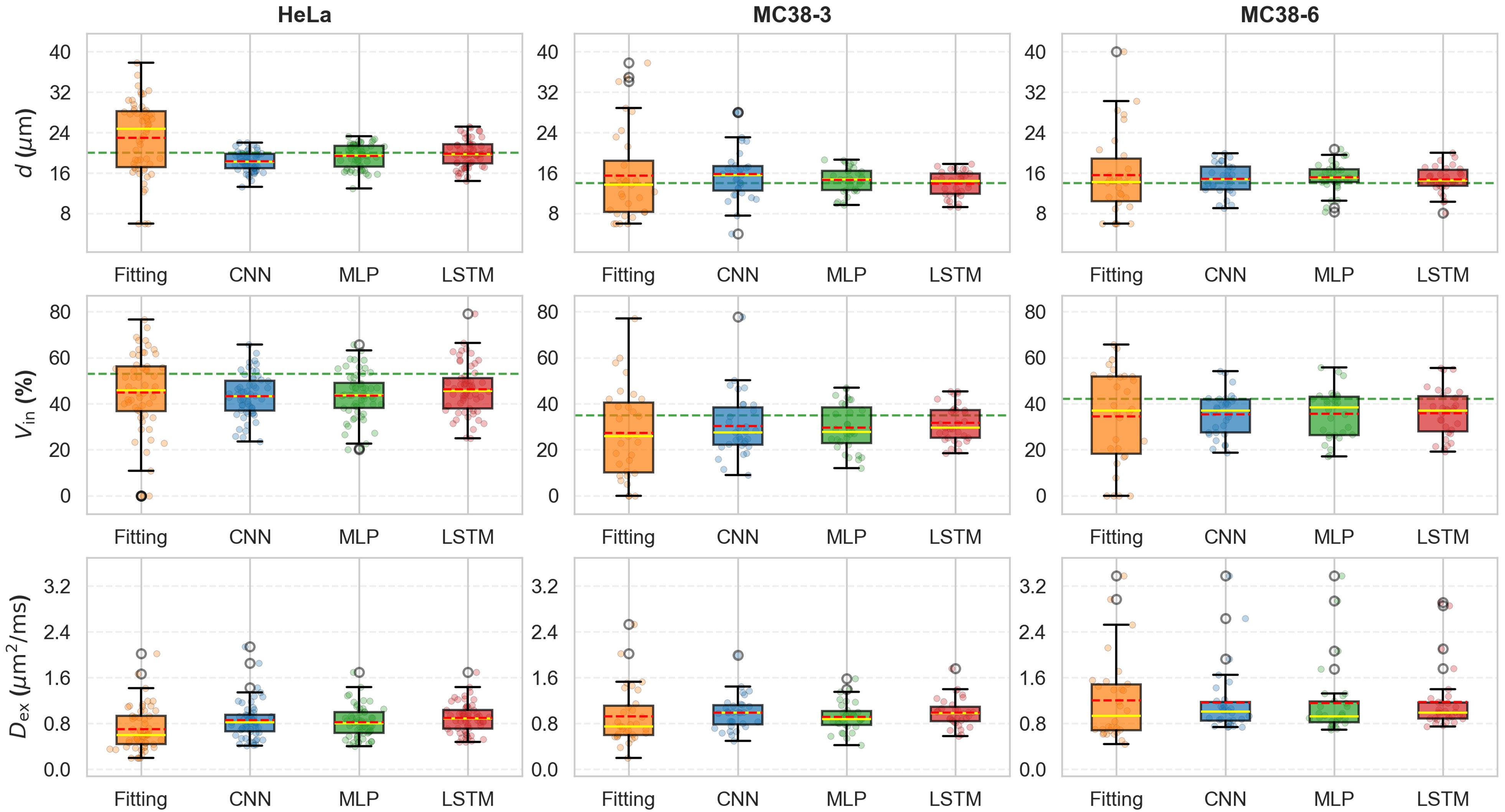}
  \caption{Comparison of four methods (Fitting, CNN, MLP, and LSTM) for estimating the cell microenvironment parameters ($d$, $V_\mathrm{in}$, and $D_\mathrm{ex}$) across three datasets (HeLa, MC38-3, and MC38-6). In each box plot, the box represents the interquartile range (IQR, 25th--75th percentiles), the solid yellow line indicates the median, and the mean is shown as dashed red line. The whiskers extend to the most extreme non-outlier values within 1.5 times the IQR, while points beyond the whiskers indicate outliers. Semi-transparent overlaid points represent randomly sampled individual fitting results, with up to 100 points displayed for each method to visualize the distribution. The green dashed lines in the $d$ and $V_\mathrm{in}$ subfigures indicate the experimentally derived reference values. No reference $D_\mathrm{ex}$ measurement was available; therefore, $D_\mathrm{ex}$ values are shown for comparison only.
}
  \label{fig:in-vitro_plot}
\end{figure}

Based on our experimental measurements, the reference cell diameter ($d$) and intracellular volume fraction ($V_\mathrm{in}$) were determined to be 20.0~$\mu$m / 53.0\% for the HeLa line, 14.0~$\mu$m / 35.0\% for MC38-3, and 14.0~$\mu$m / 42.0\% for MC38-6. As summarized in Figure~\ref{fig:in-vitro_plot}, the integration of DL-based denoising improved the agreement between the estimated microenvironment parameters and these reference values compared to direct fitting of noisy signals. 

Among the evaluated architectures, the LSTM achieved the highest estimation fidelity relative to the experimental reference values. For the HeLa cell line, the LSTM yielded an estimated $d$ of 19.8~$\mu$m and $V_\mathrm{in}$ of 46.3\%, resulting in differences of only 0.2~$\mu$m (0.8\%) and 6.7\%, respectively. Similar performance was observed for the MC38 cell lines: for MC38-3, the LSTM achieved an estimated $d$ of 13.9~$\mu$m (1.0\% difference) and $V_\mathrm{in}$ of 31.6\% (3.4\% difference); for MC38-6, the estimates were 14.8~$\mu$m (5.6\% difference) and 36.0\% (6.1\% difference). 

Consistent with the simulation results, however, the clearest experimental benefit was not only the shift in mean estimates but the improved fitting robustness. For direct fitting of raw signals, the parameter-specific failure rates were 12.9\%, 9.7\%, and 3.2\% for $d$, $V_\mathrm{in}$, and $D_\mathrm{ex}$ in MC38-3, 20.0\%, 13.3\%, and 3.3\% in MC38-6, and 5.3\% for all three parameters in HeLa, whereas all parameter-specific failure rates were reduced to 0\% by the LSTM. The overall fitting failure rate, defined as the percentage of fits in which at least one parameter reached its boundary, decreased from 19.4\%, 23.3\%, and 10.5\% to 0\% for MC38-3, MC38-6, and HeLa, respectively. These results demonstrate that physics-assisted denoising, particularly via the LSTM model, can transfer improved signal recovery into modestly improved parameter accuracy and substantially more stable biophysical fitting.

\section{Discussion}

\subsection{Low-SNR dMRI and Fitting Stability}

dMRI is a promising MRI modality for estimating cell microenvironment parameters due to its capability of probing the microscopic random motion of water molecules as they interact with cellular structures. Restrictions imposed by cell membranes and intracellular organelles modulate the diffusion signal in characteristic ways, enabling noninvasive quantification of key biophysical properties such as cell size, intracellular volume fraction, and diffusivity. However, cell parameter estimation is highly sensitive to diffusion signal quality, and imaging noise is a major challenge in low-SNR dMRI settings. This limitation becomes even more pronounced at high diffusion weightings, where the dMRI signal undergoes substantial attenuation and the relative contribution of noise increases, ultimately degrading parameter estimation accuracy and increasing the likelihood of unstable boundary-driven fits. In the context of radiotherapy, dMRI could offer an opportunity for repeated, noninvasive monitoring of tumor microenvironmental changes throughout the course of fractionated treatment, where this assessment frequency is not feasible with biopsy. Many previous dMRI-based cell microstructure studies have been conducted on 3 T or higher-field scanners \citep{jiang2016quantification,jiang2025joint,xu2021mri}, where higher SNR can support more accurate parameter estimation. However, clinical 1.5 T MRI systems may present lower-SNR conditions for quantitative dMRI, particularly when scan time is limited. Motivated by this gap, our study investigates whether denoising can improve the stability and reliability of cell microenvironment parameter fitting in low-SNR dMRI settings, with the potential to facilitate future applications on clinical 1.5 T scanners.

\subsection{DL-based Denoising Results}

The MLP and LSTM models achieved similar denoising performance, with the LSTM performing slightly better overall, whereas the CNN showed higher residual errors. The close performance of the MLP and LSTM suggests that the task may be approaching the upper bound of achievable performance under the given noise conditions, possibly because the underlying 1D denoising problem is relatively straightforward and both architectures were sufficient to reconstruct the noise-free signal from its noisy counterpart. Although we introduced Rician noise with varying levels (0.01-0.12), corresponding to SNRs ranging from 100 to 8, to represent a broad range of image-quality conditions, the denoising results remained robust across this entire spectrum. The preliminary \textit{in vitro} scanner experiment further suggests that a model trained with this synthetic noise formulation can remain effective when applied to measured dMRI data under controlled experimental conditions. Moreover, our results also showed that denoising becomes increasingly challenging at higher $b$ values due to the exponential attenuation of the diffusion signal and the associated reduction in SNR, as discussed in Section \ref{sec:denoising results}. This observation has potential implications for clinical sequence design, i.e., selecting moderately lower $b$ values may help preserve sufficient SNR to support reliable denoising performance, thereby potentially improving the robustness of future clinical dMRI applications.

For the three denoising models, the CNN exhibited the lowest performance. One possible explanation is that its convolutional filters may be less well matched to the low-dimensional 1D denoising task used here. In contrast, the MLP and LSTM both effectively captured the noise-to-signal mapping, with the LSTM yielding slightly stronger overall results. This modest advantage may reflect the LSTM's ability to model sequential dependencies in the ordered signal vector, which is consistent with the monotonic decay characteristic of dMRI signals across increasing $b$-values. However, because the performance difference between the MLP and LSTM was small, this architectural interpretation should be regarded as a plausible explanation rather than a definitive mechanism.

\subsection{Cell Microenvironment Parameters Estimation Results}

After denoising, the accuracy of cell microenvironment parameter estimation improved modestly, while fitting robustness improved substantially. In the simulation study, the estimation errors for cell diameter $d$ and intracellular volume fraction $V_\mathrm{in}$ were reduced from 5.3 $\mu$m and 15.0\% to 4.0 $\mu$m and 12.0\%, respectively (Section~\ref{sec:fitting_results}). More importantly, the same denoising step sharply reduced fitting failures, indicating that improved signal recovery translated most strongly into more stable biophysical fitting. This stability is clinically important even when the absolute accuracy gain is modest, because a parameter-estimation method must remain numerically reliable across patients, lesions, scanners, and repeated imaging time points to be practical for longitudinal assessment or treatment-response monitoring. 

Using the proposed dMRI signal denoising method, cell microenvironment parameter fitting became more stable on both simulated test data and \textit{in vitro} experimental data compared with direct fitting of raw noisy dMRI signals without denoising. In particular, the LSTM-based denoising model achieved cell-diameter differences from the reference values of $\leq$1\% for HeLa and MC38-3 and 5.6\% for MC38-6, while the intracellular volume fraction $V_\mathrm{in}$ differences remained $<$7\% across all three datasets. In the absence of reference measurements, the estimated extracellular diffusion coefficients $D_\mathrm{ex}$ were 0.9 for HeLa, 1.0 for MC38-3, and 1.2 for MC38-6. These values are consistent with previously reported results, as Jiang et al. estimated an average $D_\mathrm{ex}$ of approximately 1.0 $\mu$m$^2$/ms for MC38 cells \citep{jiang2020mri,jiang2025joint}. Beyond these modest improvements in mean estimates, denoising also substantially enhanced estimation precision, as indicated by the reduced STD across all three cell parameters. Specifically, for cell diameter $d$, the STD decreased from 7.3 to 2.7~$\mu$m (HeLa), 9.0 to 2.5~$\mu$m (MC38-3), and 8.1 to 2.6~$\mu$m (MC38-6). For $V_\mathrm{in}$, the STD was reduced from approximately 20 to 10 percentage points for all three datasets. Similarly, for $D_\mathrm{ex}$, the STD decreased from 0.4 to 0.3~$\mu$m$^2$/ms (HeLa), 0.5 to 0.3~$\mu$m$^2$/ms (MC38-3), and 0.7 to 0.5~$\mu$m$^2$/ms (MC38-6). Since the final cell microenvironment parameters are obtained by averaging voxel-wise estimates, the improved stability and robustness achieved through dMRI signal denoising may help obtain more reliable parameter estimates when only a limited number of voxels are available. This capability is particularly important for small lesions that contain only a limited number of voxels in dMRI images.

In this study, we applied two centrifugation forces (3000$\times$g and 6000$\times$g) to MC38 cells to investigate pellet compaction effects and to assess whether the proposed models are capable of distinguishing changes in the cellular microenvironment. In theory, increasing the centrifugation force from 3000$\times$g to 6000$\times$g is not expected to induce intrinsic biological changes in MC38 cell morphology, such as alterations in cell size. Instead, the primary effect of increased centrifugation is a change in cell packing density. Higher centrifugation forces result in tighter cell packing, thereby reducing the volume of the extracellular space. Consequently, an increase in the intracellular volume fraction ($V_\mathrm{in}$) is expected. In our experiments, direct measurements showed that the cell size remained unchanged under the two centrifugation conditions (approximately 14.0 $\mu$m), while $V_\mathrm{in}$ increased from 35.0\% to 42.0\%, consistent with theoretical expectations. 

Regarding the model-based estimations, all methods exhibited minimal variation in the estimated cell size $d$, with differences of less than 1.0 $\mu$m (range: 0.1-0.9 $\mu$m). Such small deviations are expected and acceptable, given experimental uncertainties (e.g., cell size measurement variability) and model fitting errors. In contrast, the estimated intracellular volume fraction $V_\mathrm{in}$ showed a clear and consistent increasing trend across all estimation methods, with changes ranging from 4.4\% to 7.2\%. This observation demonstrates that the proposed estimation models are sensitive to \textit{in vitro} changes in the cellular microenvironment induced by pellet compaction.


\subsection{Study limitations}

While the proposed framework demonstrates significant potential, several limitations warrant consideration. First, our current estimation addresses 1D models that process each voxel-wise signal independently. This discrete approach ignores spatial correlations; in heterogeneous tissues, simple voxel averaging could obscure specific tumor characteristics by conflating disparate cell types. Future work will extend this framework to 2D or 3D architectures to incorporate spatial information, which may improve both parameter estimation and tumor segmentation for radiotherapy planning.

Second, the signal model is currently optimized for a specific dMRI acquisition protocol. While we accounted for varying noise levels to approximate different SNR conditions, changes in $b$-value distribution, pulse timings, scanner hardware, or reconstruction methods may alter signal characteristics. The strong fitting stability observed in the \textit{in vitro} experiments may reflect the controlled experimental setting and favorable image quality, and should therefore be interpreted cautiously relative to the simulation results. Further validation across diverse MRI systems and imaging protocols is required to determine the generalizability and clinical robustness of the proposed method.

Finally, our \textit{in vitro} evaluation was restricted to single-cell populations (HeLa and MC38) and controlled microenvironmental changes induced by centrifugation. In contrast, the clinical tumor microenvironment is a complex milieu of tumor, immune, stromal, and vascular cells. Although our results suggest that denoising can map noisy measurements toward the noise-free IMPULSED signal patterns learned from synthetic training data, the efficacy of this approach in mixed-cell populations or \textit{in vivo} settings remains to be validated. 

\section{Conclusion}
In conclusion, this study proposed a DL-based framework for denoising dMRI signals using three architectures (CNN, MLP, and LSTM) and subsequently leveraging the denoised signals for cell microenvironment parameter estimation. DL-based processing improved dMRI signal denoising, with the MLP and LSTM achieving similar performance, the LSTM performing slightly better overall, and both outperforming the CNN. When denoised signals were used for biophysical model fitting, the improvement in parameter accuracy was modest, but the improvement in fitting robustness was substantial. In both simulation and preliminary \textit{in vitro} experiments using HeLa and MC38 cell lines, denoising reduced fitting failures and produced more stable estimates compared with direct fitting of raw noisy signals. Thus, the primary value of the proposed approach is that it preserves the interpretability of IMPULSED model fitting while making the fitting process more reliable under noisy dMRI conditions. These findings suggest that the proposed framework may improve performance in low-SNR settings and potentially facilitate future applications on clinical 1.5 T scanners. Future studies will focus on extending this framework to more realistic scenarios involving mixed cell populations, spatially resolved 2D estimation, and \textit{in vivo} or patient data to further assess the generalizability of the proposed method.


\section*{Acknowledgment}
This study was supported in part by grants from NIH (R01EB032716, R37CA214639, R01CA227289, R01CA285379). 

\newpage

\end{document}